# A Possible Explanation of Anomalous Earth Flybys

Walter Petry
Mathematisches Institut der Universitaet Duesseldorf, D-40225 Duesseldorf
E-mail: wpetry@meduse.de
petryw@uni-duesseldorf.de

Abstract: Doppler shift observations of several spacecrafts during near Earth flybys show an unexplained frequency shift. This shift is interpreted as an unexpected velocity change called Earth flyby anomaly. A theory of non-privileged reference frames is used to study the Doppler shift in such frames which are experimentally justified by the measured dipole anisotropy of the cosmic microwave background (CMB) in the solar system. The system in which the CMB is isotropic defines the privileged reference frame. The calculated frequency shift in non-privileged reference frames may give an explanation of the anomalous Earth flybys.

## 1. Introduction

The motions of several spacecrafts are studied during near Earth flybys. The Doppler frequency shifts of spacecrafts shortly before and after their nearest motions to the Earth are measured. The received anomalous frequency shift is interpretated as unexplained velocity change of the spacecraft (see e.g. [1,2]). In this preprint a possible explanation of this anomalous Earth flyby is given. It is based on privileged and non-privileged reference frames. The measured dipole anisotropy of the CMB in the solar system gives a motion of the Sun (non-privileged frame) with fixed velocity and direction with regard to a frame (privileged frame) in which the CMB radiation is isotropic. Privileged frames in which the ligth velocity is isotropic result from the Michelson-Morley experiment. These initial systems imply the well-known Lorentz transformations leaving invariant the line-element of the pseudo-Euclidean geometry. From the mathematical point of view there exist line-elements in which the light velocity is direction dependent but the result of Michelson-Morley holds too. These reference frames are called non-privileged frames and are studied by several authors. Subsequently, the theory and the results of the paper of Petry [3] about privileged and non-privileged frames are used. The Doppler frequency formula in the non-privileged reference frame of paper [3] indicates the possibility of an anomalous Earth flyby of spacecrafts. The lack of experimental results of the velocity directions of the spacecraft shortly before and shortly after the nearest Earth flyby with regard to the direction of the CMB dipole anisotopy does not allow to apply the formula for the calculated Doppler frequency anomaly. But estimates may be in agreement with the observed anomalous flybys.

## 2. Privileged and Non-Privileged Frames

In this section some results of paper [3] are stated. There exists a privileged reference frame $\Sigma'$ in which the pseudo-Euclidean geometry holds, i.e.

$$(ds')^2 = -\eta_{ij}' dx^{i'} dx^{j'} \qquad (2.1)$$

with

$$\eta_{ij}' = \delta_{ij}, \qquad i,j = 1,2,3$$
$$\eta_{i4}' = \eta_{4i}' = 0, \qquad i = 1,2,3$$
$$\eta_{44}' = -1. \qquad (2.2)$$

Here, $x^{i'}$ ($i=1,2,3$) are the Cartesian coordinates, $x^{4'} = ct'$ with the vector $x' = (x^{1'}, x^{2'}, x^{3'})$. The well-known Lorentz transformations do not change the line-element (2.1) with (2.2). In addition to the privileged frame let us consider a non-privileged frame $\Sigma$ with the Cartesian coordinates $x^i$ ($i=1,2,3$), $x^4 = ct$ and the vector $x = (x^1, x^2, x^3)$. Assume that the frame $\Sigma$ is moving with velocity $-v' = -(v^{1'}, v^{2'}, v^{3'})$ relative to the frame $\Sigma'$. Then, the line-element in $\Sigma$ is given by

$$(ds)^2 = -\eta_{ij} dx^i dx^j \qquad (2.3)$$

with

$$\eta_{ij} = \delta_{ij} - \frac{v^{i'}}{c}\frac{v^{j'}}{c}, \qquad i,j = 1,2,3$$
$$\eta_{i4} = \eta_{4i} = -\frac{v^{i'}}{c}, \qquad i = 1,2,3$$
$$\eta_{44} = -1. \qquad (2.4)$$

The relations (2.3) and (2.4) give with $ds = 0$ the absolute value of the light velocity in the frame $\Sigma$:

$$|v_l| = c / \left(1 - \left|\frac{v'}{c}\right| \cos\alpha\right). \tag{2.5}$$

Here, $|\ |$ denotes the Euclidean norm in $R^3$ and $\alpha$ is the angle between the velocity $v'$ and the light velocity $v_l$. Relation (2.5) immediately gives the null result of the Michelson-Morley experiment. Let us now consider an event in $\Sigma'$ with velocity

$$\frac{dx'}{dt'} = \left(\frac{dx^{1'}}{dt'}, \frac{dx^{2'}}{dt'}, \frac{dx^{3'}}{dt'}\right)$$

and consider the corresponding event in $\Sigma$ (i.e. not this event considered in $\Sigma$) with velocity

$$\frac{dx}{dt} = \left(\frac{dx^1}{dt}, \frac{dx^2}{dt}, \frac{dx^3}{dt}\right)$$

then it holds

$$\frac{dx}{dt} = \frac{dx'}{dt'} / \left(1 - \frac{1}{c}\left(\frac{dx'}{dt'}, \frac{v'}{c}\right)\right) \tag{2.6}$$

where $(\cdot,\cdot)$ denotes the scalar product in $R^3$. Transformation formulas from $\Sigma'$ to $\Sigma$ and conversely as well as the generalized transformation formulas of Lorentz which do not change the line-element (2.3) with (2.4) of $\Sigma$ can be found in paper [3]. Furthermore, Maxwell's equations and the equations of motion of a charged particle in a electro-magnetic field in the frame $\Sigma$ are derived in paper [3]. The experiments of Hoek and Fizeau give in the non-privileged frame $\Sigma$ the same results as in the privileged frame $\Sigma'$, i.e. they agree with the results of special relativity. The Doppler effect in the non-privileged frame $\Sigma$ is also considered in paper [3]. Assume that the motion of the frame $\Sigma$ relative to $\Sigma'$ is given by the constant velocity $-v' = -(v^{1'}, v^{2'}, 0)$. An atom moving in the frame $\Sigma$ with the constant velocity $w' = (w^{1'}, 0, 0)$ emits a photon. The frequency $\tilde{v}$ of the transverse Doppler effect, i.e. the atom moves perpendicular to the direction of the emitted light ray, is given by

$$\tilde{v} = v \left(1 - \left(\frac{w^{1'}}{c}\right)^2\right)^{1/2} / \left(1 - \frac{v^{1'}}{c}\frac{w^{1'}}{c}\right) \tag{2.7}$$

Here, $v$ is the frequency emitted by the same atom at rest in $\Sigma$. The more general case of the non-transverse Doppler effect can also be found in paper [3]. The result (2.7) can be rewritten for any constant velocities $v' = (v^{1'}, v^{2'}, v^{3'})$ and $w' = (w^{1'}, w^{2'}, w^{3'})$ in the form

$$\tilde{v} = v \left(1 - \left|\frac{w'}{c}\right|^2\right)^{1/2} / \left(1 - \left(\frac{v'}{c}, \frac{w'}{c}\right)\right). \tag{2.8}$$

It is worth mentioning that the frequency formula (2.8) is in agreement with formula (2.6) where the velocity of an object in the non-privileged reference frame is compared with the corresponding one in the privileged frame. Non-privileged reference frames in the universe are studied in the papers [4,5].

### 3. Earth Flybys Anomaly

Doppler shift data of several spacecrafts shortly before and shortly after the nearest approach to the Earth are analysed. The measured Doppler shifts can not be explained without a small but significant velocity increase of some $mm/s$ for some spacecrafts but there exist also spacecrafts with a small velocity decrease or with no velocity change. By virtue of the study of the spacecrafts near the closest approach to the Earth formula (2.8) for the transverse Doppler effect can be used. Formula (2.8) can be rewritten:

$$\tilde{v} = v \left(1 - \left|\frac{w'}{c}\right|^2\right)^{1/2} / \left(1 - \left|\frac{v'}{c}\right|\left|\frac{w'}{c}\right| \cos\vartheta\right). \tag{3.1}$$

Here, $\vartheta$ denotes the angle between the velocity $w'$ of the spacecraft and the negative velocity of the non-privileged frame, i.e. the Earth relative to the privileged frame $\Sigma'$ in which the background radiation is isotropic. The absolute value of the velocity of the Sun with regard to the privileged frame is approximately given by

$$|v'| \approx 370 km/s \tag{3.2a}$$

with the direction of the Sun-CMB:

$$(l'',b'') = (276° \pm 3°, 30° \pm 3°) \qquad (3.2b)$$

(see Kogut et al. [6]).

Let $\vartheta_b$ denote the angle shortly before the closest approach and $\vartheta_a$ the one shortly after the closest approach to the Earth then formula (3.1) gives the approximated frequency shift

$$(\tilde{\nu}_a - \tilde{\nu}_b)/\nu \approx \left|\frac{v'}{c}\right|\left|\frac{w'}{c}\right|(\cos\vartheta_a - \cos\vartheta_b). \qquad (3.3)$$

Hence, formula (3.3) yields an anomalous flyby of the spacecrafts by virtue of the non-privileged reference frame of the Earth. It can be positive, negative or zero. Formula (3.3) gives a large frequency shift if the direction of the CMB anisotropy is contained in the plane of the trajectory of the spacecraft. The frequency shift is zero if the trajectory of the spacecraft is perpendicular to the direction of the CMB anisotropy. Formula (3.3) implies the anomalous velocity change

$$\Delta v = |w'|\left|\frac{v'}{c}\right|(\cos\vartheta_a - \cos\vartheta_b). \qquad (3.4)$$

This gives for a spacecraft with velocity $|w'| \approx 13 km/s$ the result

$$\Delta v \approx 16033(\cos\vartheta_a - \cos\vartheta_b) mm/s$$

implying for a velocity increase of about $13 mm/s$ for NEAR that

$$\cos\vartheta_a - \cos\vartheta_b \approx 8 \cdot 10^{-4}. \qquad (3.5)$$

Hence, the change of the angle $\vartheta$ during the motion of the spacecraft in the neighborhood of the nearest approach is very small. This is not surprising since the angle from the center of the Earth to the spacecraft shortly before and shortly after the closest approach is already small. In addition, the direction of the CMB anisotropy with regard to the motion of the spacecraft is important. The problem of the applicability of formula (3.3) is the knowledge of the angles $\vartheta_a$ and $\vartheta_b$.

Other explanations of the Earth flybys anomaly can be found in the papers [7-10].

Summarizing, the anomalous Earth flybys of some spacecrafts is explained as anomalous frequency shift in a non-privileged reference frame, namely that of the Earth.

It is worth mentioning that in a previous paper [11] the anomalous acceleration of the Pioneers has been explained as frequency shift by considering the motion of the spacecrafts in the universe.

**References**

[ 1 ] C.Lämmerzahl, O.Preuss, and H.Dittus, Is the physics within the solar system really understood? arXiv: gr-qc/0604052
[ 2 ] J.D.Anderson, J.K.Campbell, and M.M.Nieto, The energy transfer process in planetary flybys. arXiv: astro-ph/0608087
[ 3 ] W.Petry, Does the principle of special relativity really hold? Astrophys.& Space Sci.**124** (1986),63.
[ 4 ] W.Petry, Anisotropic cosmological model and non-privileged reference frame, in: Physical Interpretations of Relativity Theory VII, London, Sept. 2000, 279.
[ 5 ] W.Petry, A theory of gravitation in flat space-time, with applications.in: Recent Advances in Relativity Theory,Vol.II, 2001, 196.
[ 6 ] A.Kogut et al., Dipole anisotropy in the COBE differential microwave radiometers. First-year sky maps. Ap.J. **419** (1993), 1.
[ 7 ] H.J. Busack, Simulation of the flyby anomaly by means of an empirical asymmetric gravitational potential with definite spatial orientation. arXiv: 0711.2781
[ 8 ] R.T. Cahill, Resolving spacecraft Earth flyby anomalies with measured light speed anisotropy. arXiv: 0804.0039
[ 9 ] M.E.McCulloch, Can the flyby anomalies be explained by a modification of inertia? arXiv: 0712.3022
[10] S.L.Adler, Can the flyby anomaly be attributed to earth-bound dark matter? arXiv: 0805.2895
[11] W.Petry, Does the Pioneer anomalous acceleration really exist? arXiv: physics/0509173